**Large Out-of-plane Polarization in Two-dimensional $Bi_2O_3$ Monolayers Predicted by an AI-driven Inverse-design Workflow**

Hongmei Tan[1,3], Jing Zhang[1], Shan Gao[1], Yueyu Zhang[2,3†], and Xiangmei Duan[1,‡]

[1]*School of Physical Science and Technology, Ningbo University, Ningbo 315211, China*

[2]*School of Integrated Circuits, Shanghai Dianji University, Shanghai 201306, China*

[3] *Wenzhou Institute, University of Chinese Academy of Sciences, Wenzhou 325001, China*

[†]Corresponding author. E-mails: yueyu.zhang@sdju.edu.cn; duanxiangmei@nbu.edu.cn

## ABSTRACT

Two-dimensional (2D) polar oxides with built-in electric fields offer promising opportunities for nanoscale electronics, optoelectronics, and surface charge engineering. However, their discovery remains challenging due to the enormous structural complexity of nonlayered oxide networks. Here, we combine inverse structural design with a machine-learning interatomic potential and first-principles calculations to explore 2D oxide configurations. Using $Bi_2O_3$ as a representative system, we identify two previously unreported polar monolayers with tetragonal and orthorhombic symmetries. Both structures exhibit vertically separated Bi and O sublayers, giving rise to pronounced intrinsic polarity with Berry-phase 2D polarizations of 34.12 and 37.15 pC/m, respectively. Notably, the o-$Bi_2O_3$ monolayer exhibits one of the largest reported out-of-plane polarizations among 2D oxides. Both monolayers exhibit dynamical stability, finite-temperature structural integrity, and wide band gaps (~3.8 eV). The large polarization originates from dimensionality-induced reconstruction of the nonlayered Bi–O network, which breaks local coordination symmetry and produces out-of-plane charge redistribution, while the stereochemically active Bi $6s^2$ lone pair further favors asymmetric Bi–O bonding. These findings suggest dimensionality-induced local-coordination symmetry breaking as a promising design motif for high-polarization 2D oxides, and demonstrate the utility of combining inverse design with machine-learning-assisted screening to explore unconventional polar phases in nonlayered oxide networks.

## I. INTRODUCTION

Intrinsic polar 2D materials have attracted increasing attention because their built-in electric fields and asymmetric surface potentials provide new opportunities for nanoscale electronics, optoelectronics, catalysis, and interface engineering[1]-[3]. In particular, out-of-plane polar 2D systems support intrinsic dipole moments across atomic thickness, enabling surface-selective charge redistribution, tunable energy-level alignment, and directional interfacial responses. Representative out-of-plane polar 2D materials span several structural classes. Van der Waals ferroelectrics such as α-$In_2Se_3$[4]-[6] and $CuInP_2S_6$[1],[7] exhibit switchable polarization in the atomically thin limit, whereas Janus monolayers such as MoSSe and WSSe possess intrinsic vertical dipoles arising from chemically inequivalent chalcogen surfaces[8]-[10]. More recently, buckled oxide monolayers have been predicted to exhibit comparatively large out-of-plane spontaneous polarizations, reaching 22.37, 23.54, and 37.58 pC/m for GeO, SnO, and PbO, respectively[11]. These examples highlight the diverse structural origins of out-of-plane polarity in two dimensions.

These systems demonstrate that surface inequivalence is an effective route to built-in polarity in two dimensions. However, most reported polar 2D materials are derived from layered or van der Waals frameworks, whereas intrinsically polar monolayers based on strongly bonded oxide networks remain largely unexplored. Whether strong out-of-plane polarity can emerge intrinsically from dimensional reduction of a nonlayered oxide network, without relying on chemically inequivalent surfaces or a pre-existing layered framework, remains unclear. Developing polar oxide monolayers is particularly attractive because metal oxides offer rich chemical flexibility, strong ionic–covalent interactions, and diverse electronic functionalities, yet their nonlayered structures make the discovery of unconventional polar phases highly challenging[12]-[14].

Bismuth oxide ($Bi_2O_3$) represents an ideal platform for exploring polar oxide monolayers due to its structural versatility and unique electronic characteristics. Bulk $Bi_2O_3$ exhibits several polymorphs, including the monoclinic α, tetragonal β, cubic γ, and cubic δ phases[15][17]. These polymorphs exhibit distinct crystal symmetries and Bi–O coordination environments, giving rise to markedly different ionic, electronic, optical, and catalytic properties[18]-[20]. The stereochemically active Bi $6s^2$ lone pair further contributes to structural flexibility and mixed ionic-covalent bonding in $Bi_2O_3$[21],[22], making $Bi_2O_3$ a particularly suitable platform for exploring low-dimensional symmetry breaking. We therefore ask whether dimensional reduction of the nonlayered Bi–O network can generate asymmetric local coordination that couples to the Bi $6s^2$ lone pair and favors an intrinsically polar geometry. Nevertheless, identifying such phases remains challenging because the strongly coupled Bi–O network contains numerous competing structural motifs that are not readily inferred from known bulk-derived prototypes. Prototype-free inverse design, such as the $IM^2ODE$ method, explores both lattice and atomic degrees of freedom without relying on predefined structural prototypes[25] and has been applied to a range of low-dimensional systems[26]-[30]. However, direct first-principles optimization of large candidate pools remains computationally demanding. Machine-learning interatomic potentials can provide rapid structural relaxation and screening before final first-principles validation[31],[32].

In the present work, we combine $IM^2ODE$-based inverse structural design with the MatterSim machine-learning interatomic potential to explore ultrathin $Bi_2O_3$ configurations. Two previously unreported $Bi_2O_3$ monolayers with tetragonal and orthorhombic symmetries, denoted t-$Bi_2O_3$ and o-$Bi_2O_3$, are identified. Both exhibit vertically separated Bi and O sublayers and large Berry-phase out-of-plane polarizations of 34.12 and 37.15 pC/m, respectively. Our analysis associates the large polarization with dimensionality-induced reconstruction of the nonlayered Bi–O network, which breaks local coordination symmetry and promotes asymmetric Bi–O bonding involving the Bi $6s^2$ lone pair. The stability and electronic, optical, transport, mechanical, and interfacial properties of the predicted monolayers are further evaluated.

## II. METHODS

### A. $IM^2ODE$-MatterSim structure search and screening

The $IM^2ODE$-MatterSim structure-search and screening workflow is summarized in Fig. 1. $IM^2ODE$ performs the population-based inverse search in the quasi-2D configurational space, while the MatterSim machine-learning interatomic potential serves as the structural optimizer within the differential-evolution iterations, enabling rapid relaxation and evaluation of newly generated candidates during the search. The resulting structures are then processed through two complementary deduplication routes: M3GNet-derived graph fingerprints identify candidates with similar local atomic environments, whereas space-group analysis removes crystallographically equivalent structures. The representatives retained from the two routes are merged into a reduced candidate set. MatterSim-based phonon calculations are subsequently used as a preliminary dynamical-stability filter, and the selected candidates are finally subjected to first-principles calculations for energetic and dynamical validation and for detailed characterization of their electronic, optical, elastic, and polar properties. This hierarchical procedure reduces structural redundancy and limits the computationally demanding DFT calculations to a small set of physically plausible candidates.

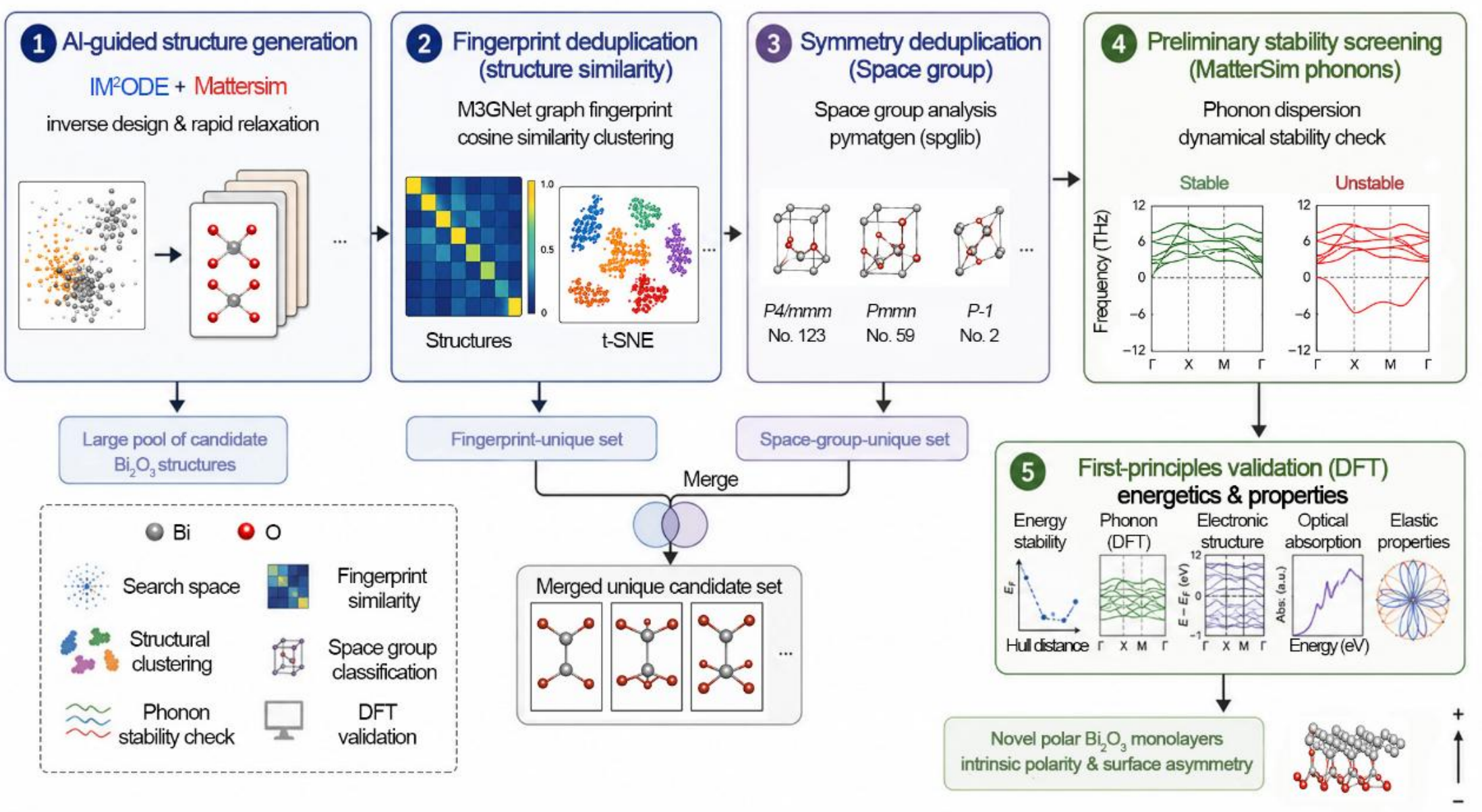

FIG. 1. Schematic of the AI-driven inverse-design workflow for ultrathin $Bi_2O_3$ monolayers, combining the coupled IM$^2$ODE-MatterSim structure search, complementary fingerprint- and symmetry-based deduplication, MatterSim-assisted stability screening, and final first-principles validation.

Two-dimensional $Bi_2O_3$ structures were explored using the IM$^2$ODE method in quasi-2D mode[25]. IM$^2$ODE employs a population-based multi-objective differential-evolution scheme to explore both lattice and atomic-coordinate degrees of freedom without relying on predefined structural prototypes. During each differential-evolution generation, the newly generated structures were relaxed using the MatterSim interatomic potential before being evaluated and propagated into the subsequent generation. Symmetry, interatomic-distance, and quasi-2D geometric constraints were imposed during structure generation, and searches were performed under different layer-thickness constraints with a 20 Å vacuum region. Using different layer-thickness constraints allows structurally distinct quasi-2D regions of the configurational space to be sampled without assuming bulk-derived structural motifs. The detailed differential-evolution operators, search parameters, and structural constraints are provided in the Supplemental Material.

A total of 800 candidate structures were obtained from the IM$^2$ODE-MatterSim search. Structural redundancy was removed using two complementary criteria. First, graph fingerprints were extracted from the final-layer node features of a pretrained M3GNet model and averaged over all atoms to obtain structure-level descriptors[31]; candidates with cosine similarities above the selected threshold were grouped as duplicates. Second, space-group analysis was performed using pymatgen/spglib to identify symmetry-equivalent candidates[33],[34]. The representative retained from the two deduplication procedures were merged into a unique candidate set, which was subsequently relaxed and ranked using the pretrained MatterSim-v1.0.0-5M interatomic potential[32]. Preliminary phonon screening based on MatterSim forces was then performed, and two structurally distinct candidates without imaginary phonon modes were selected for first-principles validation. Detailed deduplication thresholds and screening parameters are provided in the Supplemental Material.

### B. First-principles calculations

First-principles calculations were performed within density functional theory using the Vienna *Ab initio* Simulation Package (VASP)[35]. The electron-ion interactions were described using the projector-augmented-wave method, and the PBE generalized-gradient approximation was employed for structural optimization and ground-state calculations. A plane-wave cutoff energy of 500 eV was used throughout. A vacuum region larger than 20 Å was introduced along the out-of-plane direction to suppress spurious interactions between periodic images.

The in-plane lattice vectors and internal atomic coordinates were fully optimized, while the out-of-plane lattice vector was held fixed. The electronic convergence threshold was set to $10^{-8}$ eV, and structural relaxation was continued until the residual forces fell below 0.02 eV/Å. Monkhorst-Pack k-point meshes of 3×3×1 and 2×3×1 were adopted for t-$Bi_2O_3$ and o-$Bi_2O_3$, respectively.

The HSE06 screened hybrid functional, incorporating 25% short-range exact exchange with a screening parameter of 0.2 $Å^{-1}$, was employed for band structures, carrier effective masses, strain-dependent band edges, and optical responses. All calculations were performed without explicitly including spin-orbit coupling. For polar slab systems, dipole corrections were applied along the out-of-plane direction, and planar-averaged electrostatic potentials were used for dipole analysis, vacuum-level alignment, dipole-derived polarization calculations, and surface work-function calculations.

C. Polarization, phonon, and finite-temperature calculations

The out-of-plane spontaneous polarization was evaluated using the Berry-phase formulation of the modern theory of polarization[36],[37] as implemented in VASP. For each $Bi_2O_3$ phase, a linear interpolation path connecting two symmetry-related configurations with opposite out-of-plane polarizations was constructed. The polarization branch was continuously tracked along the insulating path, and the spontaneous 2D polarization was obtained from the branch-unwrapped difference between the two opposite-polarization endpoints. The interpolation path was verified to remain insulating, and dipole-derived polarizations were additionally evaluated as an independent consistency check. Detailed interpolation, polarization-branch tracking, and polarization definitions are provided in the Supplemental Material.

Phonon dispersions were calculated within density-functional perturbation theory and processed using Phonopy[38]. Finite-temperature structural integrity was examined through NVT AIMD simulations at 300, 500, and 700 K for 10 ps with a time step of 0.5 fs. Further details regarding the energetic and polarization calculations, molecular dynamics, elastic properties, carrier transport, optical response, and heterostructure construction are provided in the Supplemental Material.

## III. RESULTS AND DISCUSSION

### A. Structural properties of the predicted phases

Two distinct ultrathin $Bi_2O_3$ polymorphs—t-$Bi_2O_3$ (tetragonal) and o-$Bi_2O_3$ (orthorhombic)—were identified through the coupled IM$^2$ODE-MatterSim inverse-design search. The optimized atomic structures (Fig. 2a,b) reveal a square-ring network in t-$Bi_2O_3$ (*P*4*mm*, a=b=8.65 Å) and an anisotropic zigzag topology in o-$Bi_2O_3$ (*Pnn*2, a=10.86 Å, b=6.21 Å). Both phases exhibit $BiO_3$ coordination, with narrowly distributed Bi–O bond lengths (t-$Bi_2O_3$: 2.0905–2.0915 Å; o-$Bi_2O_3$: 2.0801–2.0931 Å; Table S1), indicating that the reduced symmetry of o-$Bi_2O_3$ arises primarily from anisotropic connectivity rather than pronounced bond-length distortion.

Side views (Fig. 2a,b) reveal vertical separation of the Bi and O sublayers, with interplanar separations of 1.00 and 1.02 Å for t-$Bi_2O_3$ and o- $Bi_2O_3$, respectively, thereby breaking inversion symmetry and generating intrinsic out-of-plane polarity. Berry-phase calculations yield spontaneous 2D polarizations $|P_s^{2D}|$ of 34.12 and 37.15 pC/m for t-$Bi_2O_3$ and o-$Bi_2O_3$, respectively (Table I). The corresponding interpolation paths remain insulating, enabling continuous tracking of the polarization branches (Fig. S1). These values are among the largest reported out-of-plane polarizations of representative monolayers, with o-$Bi_2O_3$ only 0.43 pC/m below the reported PbO value[11].

Electron localization function (ELF)[39] distributions (Fig. 2c,d) reveal pronounced O-centered localization, originating from the O-2p-derived valence electrons and consistent with oxygen's higher electronegativity relative to Bi. The elongated, nonspherical ELF basins reflect the anisotropic local coordination and directional character of the Bi–O interactions, while finite ELF along the Bi–O linkages indicates a non-negligible covalent contribution to the strongly polar bonds. The asymmetric coordination around Bi is also consistent with the stereochemical activity of the Bi 6$s^2$ lone pair[21],[22]. Taken together, the large out-of-plane polarization is associated with dimensionality-induced reconstruction of the nonlayered Bi–O network, which breaks local coordination symmetry and produces an out-of-plane charge redistribution, while the stereochemically active Bi $6s^2$ lone pair favors asymmetric Bi–O bonding.

TABLE I Comparison of reported out-of-plane 2D polarization magnitudes $|P_z^{2D}|$ in representative polar and ferroelectric monolayers.

| **Material** | $|P_z^{2D}|$ **(pC/m)** |
|---|---|
| **2D oxide monolayers** | |
| PbO[11] | 37.58 |
| **o-$Bi_2O_3$ (this work)** | **37.15** |
| **t- $Bi_2O_3$ (this work)** | **34.12** |
| SnO[11] | 23.54 |
| GeO[11] | 22.37 |
| h-PbO[40] | ≈28 |
| **vdW ferroelectric monolayers** | |
| $In_2Se_3$[1] | 16.9 |
| $CuInP_2S_6$[1] | 4.99 |
| $CuInP_2Se_6$[1] | 3.67 |
| $CuBiP_2Se_6$[1] | 2.51 |
| $AgBiP_2Se_6$[1] | 1.39 |
| **Other polar monolayers** | |
| $CSiH_2$[1] | 19.4 |
| $CSiF_2$[1] | 0.656 |
| InP[1] | 12.5 |
| CdTe[41] | 9.27 |
| CdSe[41] | 9.13 |
| Q-GaSe[42] | 4.89 |

**Note**. Literature values are theoretical out-of-plane 2D polarizations reported in the cited studies. The values for o-$Bi_2O_3$ and t-$Bi_2O_3$ are Berry-phase spontaneous polarizations obtained in this work. Different computational protocols may lead to quantitative variations among the reported values.

### B. Thermodynamic, dynamical, and finite-temperature stability

Following the characterization of structural polarity, we assessed the thermodynamic stability of the two $Bi_2O_3$ monolayers by calculating their formation energies ($E_f$) and energies above the convex hull ($E_{hull}$) within the PBE framework. As summarized in Table II, the $E_f$ values for t-$Bi_2O_3$ and o-$Bi_2O_3$ are −0.84 and −0.84 eV/atom, respectively. These negative values confirm that the formation of both monolayers is energetically favorable relative to the chosen elemental reference states. The energy difference between the two phases is negligible (~2.4 meV/atom), precluding a definitive energetic ranking at the present level of theory.

Relative to the bulk ground state, the calculated $E_{\text{hull}}$ values are 0.37 eV/atom for t-$Bi_2O_3$ and 0.37 eV/atom for o-$Bi_2O_3$. These positive values confirm that both monolayers are thermodynamically metastable phases[43], while their comparable magnitudes indicate a similar degree of metastability.

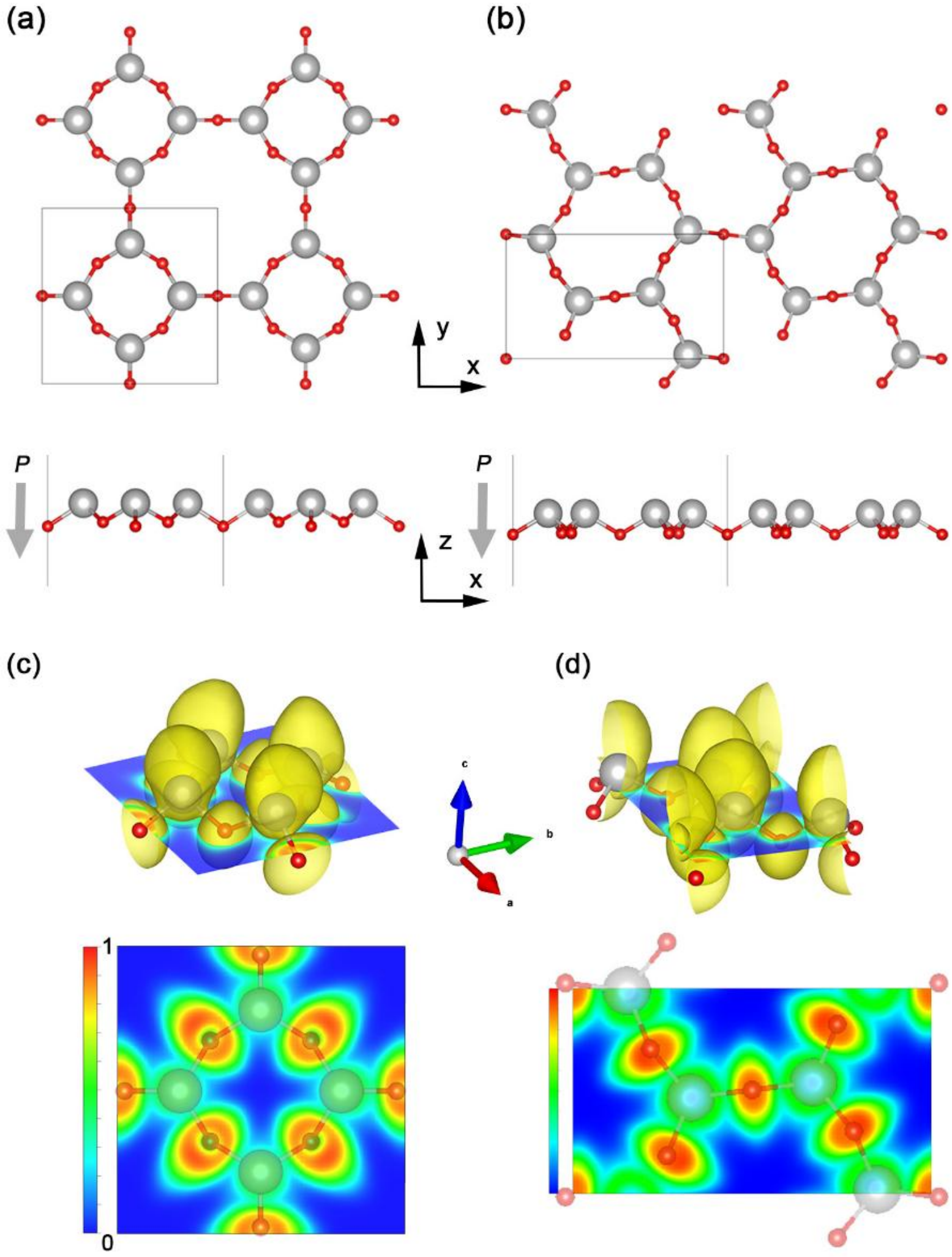

FIG. 2. Atomic structures and electron localization functions of predicted 2D $Bi_2O_3$ phases. (a, b) Top and side views of (a) t-$Bi_2O_3$ and (b) o-$Bi_2O_3$, where gray and red spheres denote Bi and O atoms, respectively. Arrows indicate the out-of-plane polarity. (c, d) 3D ELF isosurfaces (isovalue = 0.40) and corresponding 2D ELF maps on (001) planes for (c) t-$Bi_2O_3$ ($z$ = 0.52) and (d) o-$Bi_2O_3$ ($z$ = 0.50). The color bar indicates ELF values from 0 to 1.

To determine whether these energetically metastable monolayers correspond to local minima and retain their structural frameworks at finite temperatures, we examined their phonon dispersions and AIMD simulations. The phonon spectra of both t-$Bi_2O_3$ and o-$Bi_2O_3$ exhibit no significant imaginary modes (Fig. 3a, b), confirming their dynamical stability within the harmonic approximation.

TABLE II. Calculated PBE formation energy $E_\mathrm{f}$ and energy above the convex hull $E_\mathrm{hull}$, PBE and HSE06 band gaps of t-$Bi_2O_3$ and o-$Bi_2O_3$.

| Phase | $E_\mathrm{f}$ (eV/atom) | $E_\mathrm{hull}$ (eV/atom) | $E_\mathrm{g}^\mathrm{PBE}$ (eV) | $E_\mathrm{g}^\mathrm{HSE}$ (eV) |
|---|---|---|---|---|
| t-$Bi_2O_3$ | –0.84 | 0.37 | 2.30 | 3.71 |
| o-$Bi_2O_3$ | –0.84 | 0.37 | 2.32 | 3.79 |

Finite-temperature structural stability was subsequently evaluated via 10 ps AIMD simulations in the NVT ensemble. For t-$Bi_2O_3$, the total energy fluctuates around a stationary average following a brief initial equilibration at both 300 and 500 K (Fig. 3c, e). The corresponding RMSD values rapidly converge to bounded plateaus, and the structures averaged over the final 0.5 ps retain the characteristic square-ring network. These results indicate that the bonding framework of t-$Bi_2O_3$ remains intact throughout the simulations at both temperatures.

For o-$Bi_2O_3$, the total energy first decreases at 300 K and then stabilizes (Fig. 3d). At 500 K, a longer equilibration period precedes stabilized energy fluctuations (Fig. 3f). Although o-$Bi_2O_3$ exhibits larger RMSD and energy fluctuations than t-$Bi_2O_3$, the RMSD remains bounded with no Bi-O bond breaking or coordination change. Enhanced fluctuations at 700 K (Fig. S2) indicate stronger thermal perturbations, yet both phases preserve their bonding framework over 10 ps. Thus, AIMD confirms a stronger finite-temperature response in o-$Bi_2O_3$ while affirming the structural integrity of both monolayers.

### C. Electronic structures and band-edge properties

Having established the local stability of both monolayers, we proceed to analyze their electronic structures, carrier-transport properties, and optical responses. As summarized in Table 1, the PBE band gaps of t-$Bi_2O_3$ and o-$Bi_2O_3$ are 2.30 and 2.32 eV, respectively, while the corresponding HSE06 values increase to 3.71 and 3.79 eV.

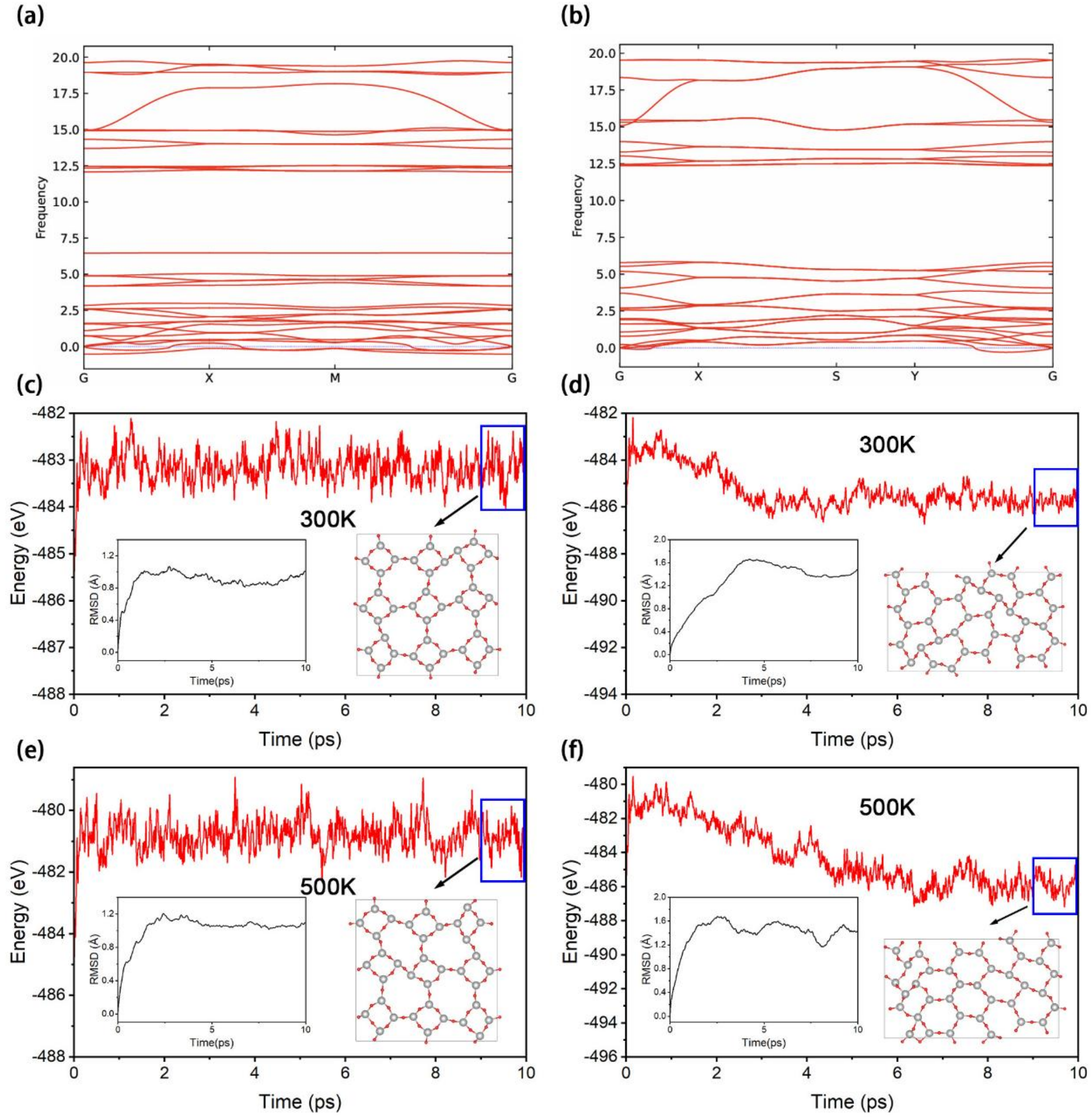

FIG. 3. Phonon dispersions and AIMD simulations of the predicted 2D $Bi_2O_3$ phases. (a, b) Phonon spectra of (a) t-$Bi_2O_3$ and (b) o-$Bi_2O_3$. (c–f) Time evolution of the total energy during NVT AIMD simulations at 300 and 500 K for t- $Bi_2O_3$ [(c, e)] and o-$Bi_2O_3$ [(d, f)], respectively. Insets display the RMSD relative to the optimized initial structures and the atomic configurations averaged over the final 0.5 ps.

Figure 4a and 4b present the HSE06 band structures. For t-$Bi_2O_3$, the valence-band maximum (VBM) is located at the M point and the conduction-band minimum (CBM) at the Γ point. For o-$Bi_2O_3$, the VBM occurs at k = (0.25, 0, 0) along the Γ–X direction, whereas the CBM remains at Γ. The minimum direct gaps are 3.89 eV for t-$Bi_2O_3$ and 3.79 eV for o-$Bi_2O_3$, yielding an indirect-to-direct gap difference of 0.18 eV for the former phase. In contrast, the indirect and direct gaps of o-$Bi_2O_3$ are indistinguishable within the reported numerical precision.

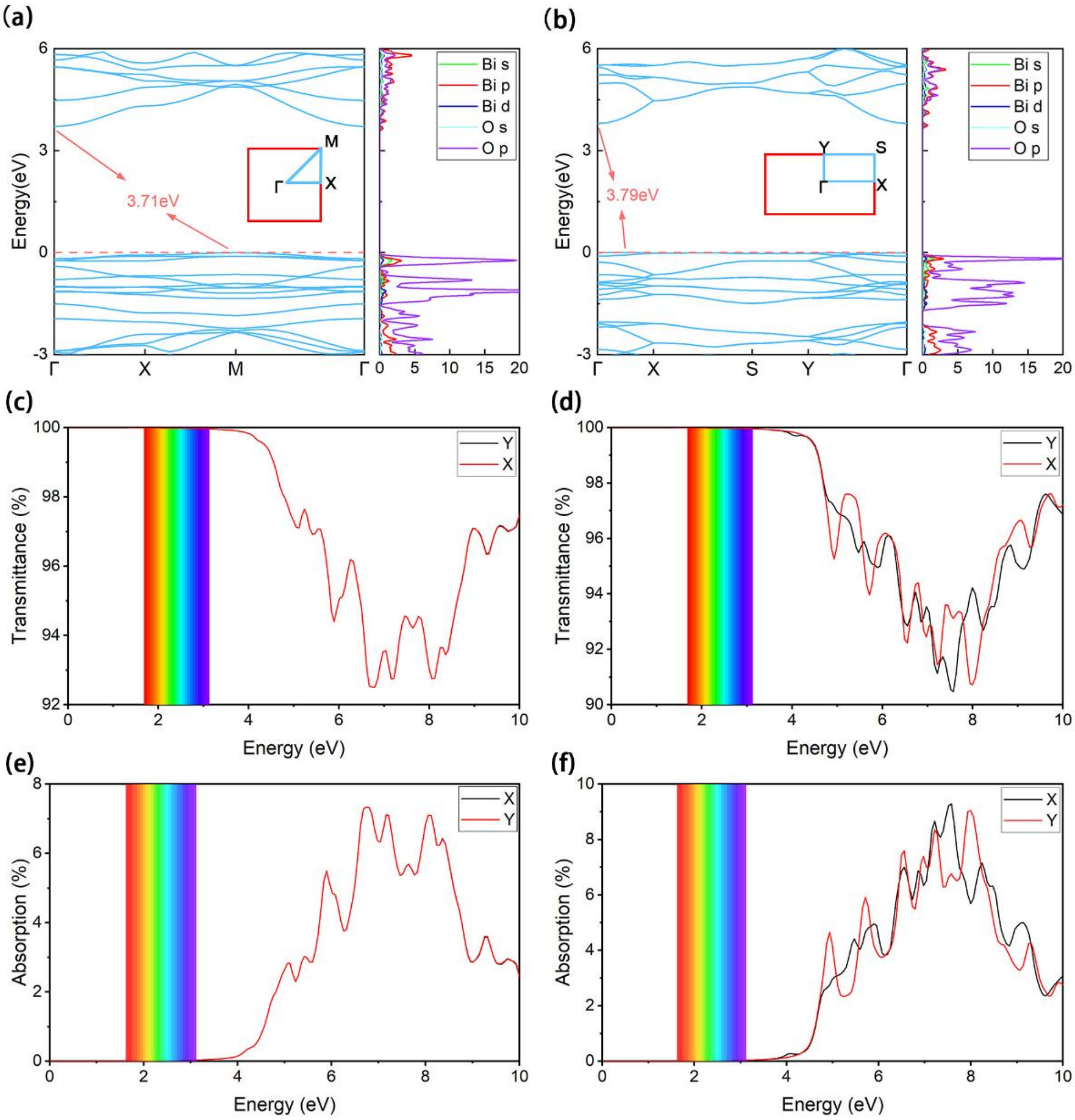

FIG. 4. (a, b) HSE06-calculated band structures and projected density of states for (a) t-$Bi_2O_3$ and (b) o-$Bi_2O_3$. The energy zero is set to the Fermi level ($E_F = 0$ eV). Insets indicate the high-symmetry k-point paths in the first Brillouin zone. (c, d) Transmittance spectra for (c) t-$Bi_2O_3$ and (d) o-$Bi_2O_3$, and (e, f) absorptance spectra for (e) t-$Bi_2O_3$ and (f) o-$Bi_2O_3$, calculated along the *x* and *y* directions. The colored regions highlight the visible light range (~1.65-3.10 eV).

The projected density of states reveals similar band-edge compositions in both phases. The VBM is dominated by O-*p* states, while the CBM primarily comprises Bi-*p* states with a minor O-*p* contribution. The relatively flat valence bands and more dispersive conduction bands are consistent with the substantially heavier holes and lighter electrons extracted from the HSE06 band-edge fitting.

## D. Carrier transport and optical response

The acoustic-phonon-limited carrier transport at 300 K, listed in Table S2, reflects the underlying symmetry of each monolayer. Owing to the fourfold in-plane symmetry of t-$Bi_2O_3$, its transport parameters are isotropic along the *x* and *y* directions. The electron effective mass is 0.456 $m_0$, considerably smaller than the hole effective mass of 3.011 $m_0$. Combined with a smaller electron deformation-potential constant, this yields an estimated electron mobility of 703.52 $cm^2$/V/s, more than two orders of magnitude greater than the hole value of 2.86 $cm^2$/V/s.

The orthorhombic o-$Bi_2O_3$ phase exhibits pronounced directional dependence. Its electron effective masses are 0.415 $m_0$ and 0.492 $m_0$ along *x* and *y*, respectively, with corresponding mobilities of 199.72 and 54.28 $cm^2$/V/s. The faster electron transport along *x* arises from the combined effects of a larger 2D elastic modulus, a smaller effective mass, and a lower deformation-potential constant. Hole transport displays the opposite directional preference, increasing from 2.54 $cm^2$/V/s along *x* to 8.52 $cm^2$/V/s along *y*. Although the hole effective mass is larger and the elastic modulus is smaller along *y*, the substantially reduced deformation-potential constant—0.624 eV compared with 1.736 eV along *x*—dominates the calculated transport rate. These results reveal carrier-dependent transport anisotropy in o-$Bi_2O_3$, with faster electron mobility along *x* and higher hole mobility along *y*.

The HSE06 optical spectra are consistent with the wide-band-gap electronic structures. As shown in Fig. 4c–4f, both monolayers exhibit nearly complete transmittance and negligible absorptance throughout the visible range (~ 1.65–3.10 eV). Optical absorption becomes appreciable near 3.8–4.0 eV, in agreement with their minimum direct gaps, and is concentrated mainly in the ultraviolet region. For t-$Bi_2O_3$, the in-plane (*x*- and *y*-polarized) spectra nearly overlap, reflecting its tetragonal symmetry. In contrast, o-$Bi_2O_3$ exhibits clear polarization-dependent differences in the ultraviolet region, consistent with its orthorhombic lattice and anisotropic electronic structure. These results indicate that the in-plane symmetry governs both carrier transport and optical anisotropy, while both monolayers combine visible-light transparency with ultraviolet absorption.

### E. Anisotropic mechanical properties

The symmetry-dependent anisotropy identified in the carrier-transport and optical responses is also reflected in the in-plane mechanical properties. The calculated elastic tensors of both t-$Bi_2O_3$ and o-$Bi_2O_3$ satisfy the corresponding 2D Born stability criteria[44], confirming their mechanical stability within the linear elastic regime. Figure 5 presents the orientation-dependent Young's modulus ($E$), shear modulus ($G$), and Poisson's ratio for each monolayer.

For t-$Bi_2O_3$, all three mechanical parameters exhibit fourfold rotational symmetry, consistent with its square lattice (Fig. 5a–c). Taking the 0° and 90° directions as $x$ and $y$, respectively, the Young's modulus reaches a maximum of 15.53 N/m along the principal axes and decreases to a minimum of 7.30 N/m along the diagonal directions at 45° and 135°, yielding an anisotropy ratio $E_{max}/E_{min}$ of 2.13. The shear modulus varies from 2.09 to 5.34 N/m, corresponding to an anisotropy ratio of 2.55, while the Poisson's ratio ranges from 0.45 to 0.74.

The orthorhombic symmetry of o-$Bi_2O_3$ gives rise to a strongly anisotropic mechanical response (Fig. 5d–f). Its Young's modulus reaches a maximum of 15.32 N/m along off-axis directions near 34.5° and 145.5°, and decreases to a minimum of 5.73 N/m along the $y$ direction, corresponding to an anisotropy ratio of 2.68. The shear modulus varies from 2.43 to 5.78 N/m, with maxima along the principal axes and minima along the diagonal directions, yielding an anisotropy ratio of 2.38. The Poisson's ratio remains positive throughout and ranges from 0.18 to 0.81, indicating pronounced direction-dependent transverse strain coupling.

These results confirm that the in-plane lattice symmetry governs not only the electronic and optical anisotropy but also the directional elastic responses of the two $Bi_2O_3$ monolayers.

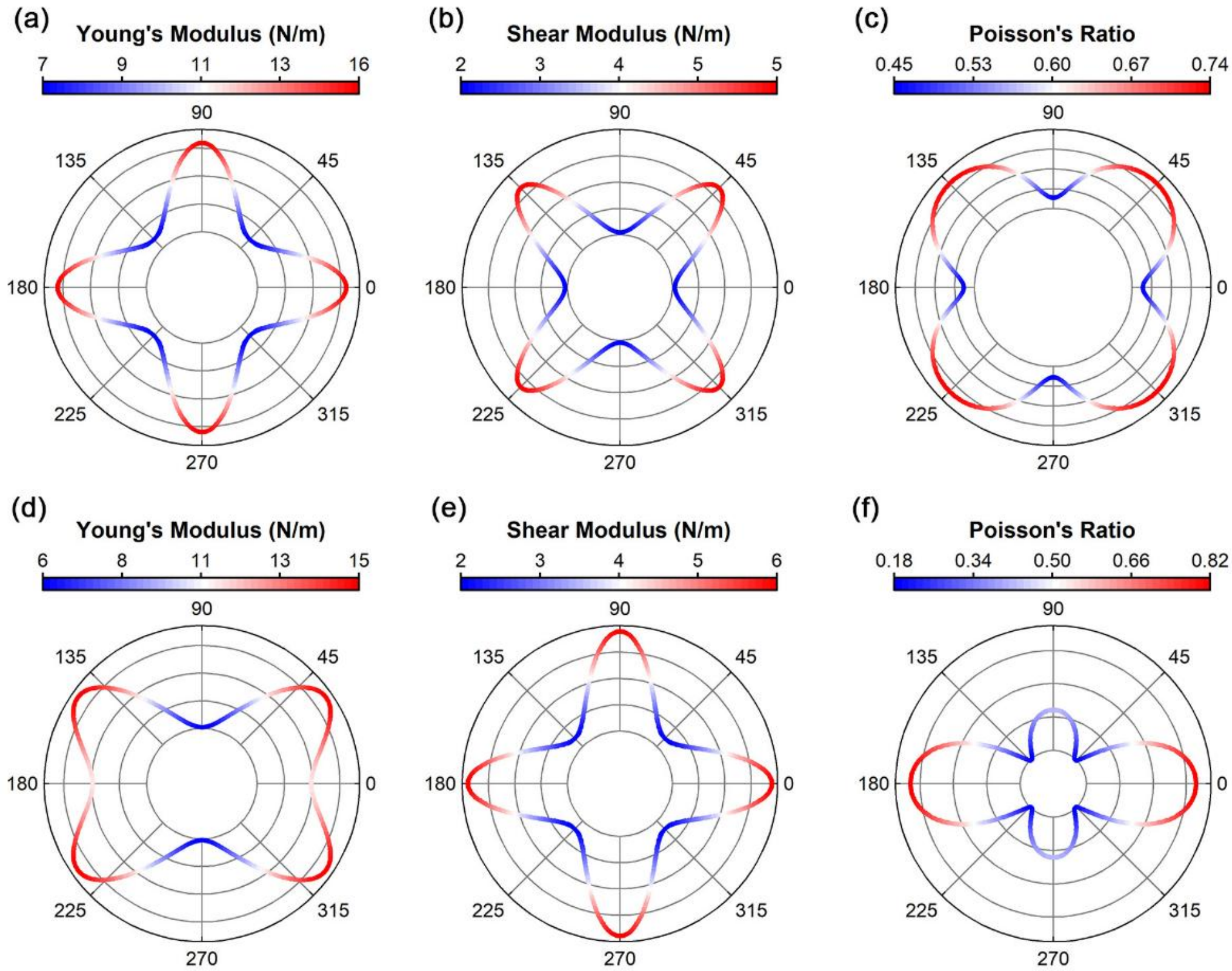

FIG. 5. Orientation-dependent in-plane mechanical properties of 2D $Bi_2O_3$ phases. (a–c) Polar plots of the Young's modulus, shear modulus, and Poisson's ratio for t-$Bi_2O_3$. (d–f) Corresponding mechanical anisotropy profiles for o-$Bi_2O_3$. The angle 0° and 90° align with the crystallographic $x$ and $y$ axes, respectively. The color gradient represents the magnitude of each mechanical parameter.

### F. Surface work-function asymmetry and its modulation in heterostructures

While the above analyses establish the symmetry-dependent in-plane anisotropy of both monolayers, the key feature that distinguishes these materials from conventional 2D oxides is their intrinsic out-of-plane polarity, which we examine next. As shown in Fig. 6a and 6b, the O-terminated and Bi-terminated surfaces of the isolated $Bi_2O_3$ monolayers exhibit markedly different vacuum levels. For t-$Bi_2O_3$, the work functions are 9.22 eV on the O-terminated side and 4.95 eV on the Bi-terminated one, corresponding to a surface contrast of 4.27 eV. The corresponding values for o-$Bi_2O_3$ are 9.37 and 4.72 eV, resulting in a slightly larger contrast of 4.65 eV. In both phases, the higher work function of the O-terminated surface reflects the electrostatic potential step arising from the vertically separated Bi and O sublayers and the resulting built-in dipole[45][45].

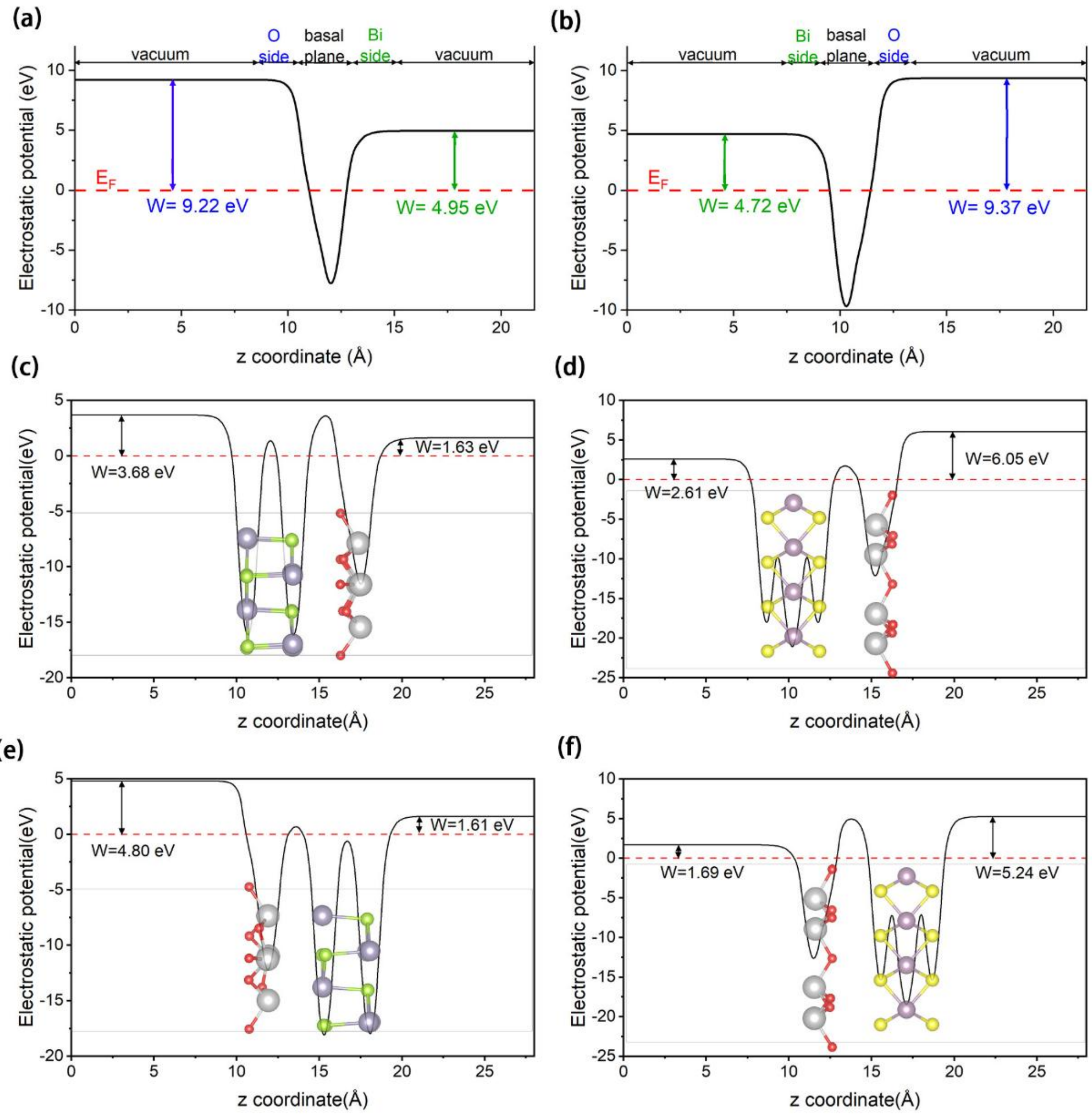

FIG. 6. Planar-averaged electrostatic potentials and surface work functions of isolated $Bi_2O_3$ monolayers and their heterostructures. Results are presented for (a) isolated t-$Bi_2O_3$, (b) isolated o-$Bi_2O_3$, (c) Bi-contact t-$Bi_2O_3$/SnSe, (d) Bi-contact o-$Bi_2O_3$/$MoS_2$, (e) O-contact t-$Bi_2O_3$/SnSe, and (f) O-contact o-$Bi_2O_3$/$MoS_2$. The terms Bi-contact and O-contact denote that heterostructures where Bi-terminated and O-terminated surfaces of $Bi_2O_3$, respectively, are positioned at the interface. The black curves, red dashed lines, and arrows indicate the planar-averaged electrostatic potentials, Fermi levels, and surface work functions, respectively.

To examine how this surface asymmetry evolves upon interface formation, relaxed t-$Bi_2O_3$/SnSe and o-$Bi_2O_3$/$MoS_2$ heterostructures were considered in two opposite orientations. Here, the Bi-contact and O-contact configurations denote that the Bi-terminated and O-terminated sides of $Bi_2O_3$ are placed at the interface, respectively.

For the Bi-contact t-$Bi_2O_3$/SnSe configuration, the outer SnSe surface and the exposed O-terminated $Bi_2O_3$ surface exhibit work functions of 3.68 and 1.63 eV, respectively, yielding a contrast of 2.05 eV (Fig. 6c). Reversing the orientation to the O-contact configuration gives work functions of 4.80 eV for the exposed Bi-terminated $Bi_2O_3$ surface and 1.61 eV for the outer SnSe surface, corresponding to a larger contrast of 3.19 eV (Fig. 6e).

For the Bi-contact o-$Bi_2O_3$/$MoS_2$ configuration, the work functions of the outer $MoS_2$ surface and the exposed O-terminated $Bi_2O_3$ surface are 2.61 and 6.05 eV, respectively, corresponding to a contrast of 3.44 eV (Fig. 6d). In the O-contact configuration, the exposed Bi-terminated $Bi_2O_3$ surface exhibits a work function of 1.69 eV, whereas that of the outer $MoS_2$ surface is 5.24 eV, yielding a similar contrast of 3.55 eV (Fig. 6f). Thus, reversing the $Bi_2O_3$ orientation markedly shifts the absolute work functions of the exposed surfaces. Its effect on the magnitude of the two-sided contrast is heterostructure dependent: The contrast changes from 2.05 to 3.19 eV in t-$Bi_2O_3$/SnSe but remains close to 3.5 eV in o-$Bi_2O_3$/$MoS_2$.

The work-function contrasts of all four heterostructures are smaller than those of the corresponding isolated $Bi_2O_3$ monolayers, although sizable two-sided asymmetry persists. This reduction is consistent with partial compensation of the intrinsic $Bi_2O_3$ dipole by interface-induced electrostatic redistribution. The orientation of the polar $Bi_2O_3$ layer therefore provides a direct means of tuning the surface energy levels and electrostatic potential distribution across the heterostructure. Collectively, these characteristics—intrinsic polarity, local stability, wide-band-gap electronic structures, symmetry-dependent transport, optical and mechanical responses, and orientation-sensitive surface potentials—position these predicted $Bi_2O_3$ monolayers as candidate polar building blocks for 2D vdW heterostructures.

## G. CONCLUSION

In summary, we combined $IM^2ODE$-based inverse structural design with the MatterSim machine-learning interatomic potential and first-principles calculations to explore unconventional ultrathin $Bi_2O_3$ configurations. Two previously unreported intrinsically polar monolayers with tetragonal and orthorhombic symmetries were identified. Both phases exhibit vertical separation of the Bi and O sublayers, which breaks out-of-plane inversion symmetry and gives rise to large Berry-phase spontaneous polarizations of 34.12 and 37.15 pC/m, respectively. The large polarization is associated with dimensionality-induced reconstruction of the nonlayered Bi–O network, which breaks local coordination symmetry and produces out-of-plane charge redistribution, while the Bi $6s^2$ lone pair favors asymmetric Bi–O bonding. Although thermodynamically metastable at approximately 0.37 eV/atom above the bulk convex hull, both monolayers exhibit dynamical stability and finite-temperature structural integrity, as supported by phonon and AIMD calculations. They also possess wide HSE06 band gaps of approximately 3.7–3.8 eV, together with symmetry-dependent carrier transport and optical responses. Their intrinsic polar asymmetry further produces work-function contrasts exceeding 4 eV between the two surfaces, which can be modulated by reversing the $Bi_2O_3$ orientation in vdW heterostructures. These results suggest dimensionality-induced local-coordination symmetry breaking as a promising route to large out-of-plane polarization in 2D oxides and demonstrate the utility of combining inverse design with machine-learning-assisted screening to explore unconventional polar phases in nonlayered oxide networks.

## ACKNOWLEDGMENTS

This work is supported by the National Natural Science Foundation of China (Grant No. 12374061, U23A2072, and 12404274), Zhejiang Provincial Natural Science Foundation of China (Grant No. LD24F040001, and LQN25A040020), and the KC Wong Magna Foundation in Ningbo University.

## CODE AND DATA AVAILABILITY

The implementation of the IM$^2$ODE-based structure-search workflow and its coupling with MatterSim for preliminary relaxation and screening is available at https://github.com/YueyuZhang/IM2ODE4mattersim. Other input files and post-processing scripts are available from the corresponding author upon reasonable request.

The optimized atomic structures, additional computational details, energetic and polarization definitions, AIMD results, elastic constants, carrier-transport parameters, optical calculation settings, and heterostructure construction details are provided in the Supplemental Material. Additional data supporting the findings of this study are available from the corresponding author upon reasonable request.

## SUPPLEMENTAL MATERIAL

Additional computational details; optimized atomic coordinates of t-$Bi_2O_3$, o-$Bi_2O_3$, elemental reference phases, and bulk $Bi_2O_3$; definitions of formation energy, convex-hull energy, and Berry-phase and dipole-derived 2D polarizations; Berry-phase polarization paths; AIMD results at 700 K; elastic-property calculations; carrier mobility parameters; optical calculation settings; and heterostructure construction details.